\documentstyle[epsfig]{mn}
\onecolumn
\input{epsf}
\def\n{{\noindent}}
\def\rbar{\bar{r}}

\title[Galaxy Clustering and Dark Energy]
{Galaxy Clustering and Dark Energy}

\author[D.Munshi et al.]
{Dipak Munshi$^{1,2}$,  
Cristiano Porciani$^{1,3}$, Yun Wang$^{4}$\\
 $^1$Institute of Astronomy, Madingley Road,
Cambridge - CB3 OHA, United Kingdom\\
$^2$Astrophysics Group, Cavendish Laboratory, Madingley Road, Cambridge CB3 OHE, United Kingdom\\
$^3$Institute of Astronomy, Dept. of Physics, ETH H\"onggerberg,
8093 Z\"urich, Switzerland\\
$^4$Dept. of Physics \& Astronomy, University of Oklohama, Norman, OK 73019, USA  \\
}

\pagerange{000,000}

\def\la{\mathrel{\mathpalette\fun <}}
\def\ga{\mathrel{\mathpalette\fun >}}
\def\fun#1#2{\lower3.6pt\vbox{\baselineskip0pt\lineskip.9pt
        \ialign{$\mathsurround=0pt#1\hfill##\hfil$\crcr#2\crcr\sim\crcr}}}
        
\begin{document}
\maketitle

\begin{abstract}
We study the evolution of galaxy clustering in various cosmological models 
with quintessence. We investigate how the analytical predictions vary with 
change of dark energy equation of state $w_X$. 
Comparing these predictions against available data we 
discuss to what extent the problems of galaxy 
biasing can be modelled.
This will be key in constraining the dark energy equation of state with 
future galaxy surveys. 
We use a compilation
of various surveys to study the number density and amplitude of 
galaxy clustering from observations 
of the local universe at $z \sim 0$ to that of the Lyman break galaxies
and Ly-$\alpha$ emitters at $z \sim 4.9$. 
We find that there is a degeneracy between the dark energy equation of state 
and the way
galaxies populate dark matter haloes;
objects are more biased in models with more negative values 
of dark energy equation of state $w_X$.

We conclude that, while 
future all sky CMB observations will determine cosmological 
parameters with unprecedented precision, and cross correlation 
of weak lensing experiments and galaxy surveys will provide a cleaner
and accurate picture of bias associated with collapsed objects, the rate of
growth of large scale structure in such surveys 
can potentially constrain the equation of state of dark energy and 
the potential of the scalar field associated with quintessence.
In particular, we show that the abundance and spatial distribution of
galaxy clusters at intermediate redshifts
strongly depend on the dark energy equation of state. 
When accurate measurement of
galaxy clustering at high-redshit becomes possible, it will 
provide constraints on dark energy
that are independent and complementary to type Ia supernova studies.
\end{abstract}


\section{Introduction}

It is generally believed that small perturbations in the matter density, generated
by quantum effects during inflation, eventually grow
due to gravitational instability, and finally collapse to produce luminous
objects such as galaxies and clusters which can be observed today.
The evolution of galaxy clustering can be used to constrain cosmological models and
the dark matter scenarios. In particular, the evolution of clustering with redshift can 
put direct constraints on models for the evolution of density perturbations. In this paper we study
how the equation of state of dark energy affects the observed
clustering of luminous objects.

For many years the study of the spatial distribution of galaxies 
at high redshift
has been rather sketchy and affected by various observational limitations.
Early studies showed that galaxy clustering,
when parameterised by the 
rms amplitude of fluctuations in the galaxy counts within a fixed comoving 
scale, typically decreases with
redshift for moderately deep samples $(0<z\la 1)$.
Recent progress in colour selection criteria has made empirical studies of
the high redshift universe possible observationally. Colour selection such as 
the Lyman-break
technique (Steidel et al. 1996, 1998; Madau et al. 1996; Lowenthal et al. 1997)
or the photometric redshift technique (for example, see 
Wang, Bahcall, \& Turner 1998,
Budavari et al. 2000, 
Fernandez-Soto et al. 2001), allows one to
efficiently identify classes of galaxies in a preassigned redshift range 
based on their spectral energy distribution. This has resulted in the 
compilation of large and well-controlled
samples of galaxies at $z>2$ which are suitable for clustering studies 
(see e.g. Porciani \& Giavalisco 2001 and references therein for details).
These studies measured a very strong clustering amplitude, 
comparable to that of present-day galaxies.
It is worth stressing, however, that
Lyman-break galaxies (LBG hereafter) essentially consist of actively 
star-forming
galaxies; in comparison, quiescent galaxies at high redshifts are much less 
efficiently identified with current instrumentation. 

Although the detection of strong clustering seems to be quite 
robust at high redshift, the current samples still contain too few objects 
and cover too small an area on the sky to accurately measure the corresponding 
correlation functions. The signal-to-noise ratio of the current measurements 
is of order 3 for the 2-point statistics,
and the dispersion among different measurements 
suggests the possibility of systematic errors (Porciani \& Giavalisco 2002). 
Robust statistical techniques combined with
next generation extensive surveys can greatly enhance our 
knowledge of clustering of high redshift galaxies, allowing 
to use it as a test for cosmological scenarios.

\begin{figure}
\protect\centerline{
\epsfysize = 5.0truein
\epsfbox[21 147 588 715]
{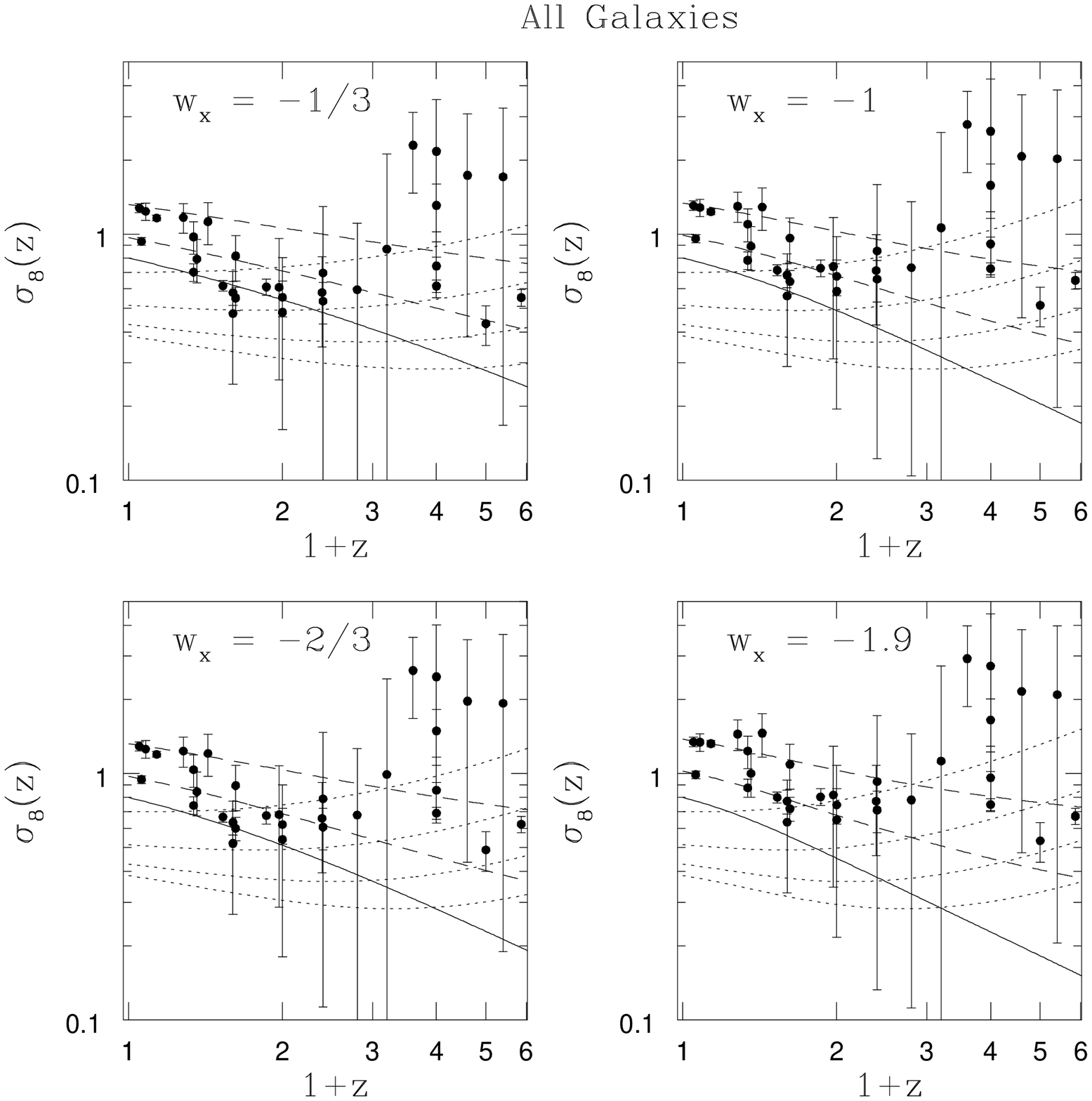} }
\caption{Analytical computations of $\sigma_8$ and $\sigma_8^{(g)}$ 
as a function of redshift $z$
 for various quintessence models are compared with observational results. The solid lines
represent the linear growth rate $D(z)$ for various cosmologies as a function of redshift.
The short-dashed lines represent the theoretical $\sigma_8(z)$; 
the two dashed lines are normalized to APM and IRAS surveys at low redshift
respectively. The dotted lines represent the predictions from halo model.
We use the analytical results of Mo \& White (1996) to compute the bias parameter for haloes larger 
than a given mass threshold. Curves from bottom upwards correspond to haloes with masses greater that 
$10^9$, $10^{10}$, $10^{11}$, $10^{12}$ M$_\odot$. The various data sets 
consist of large galaxy surveys at low redshift such as IRAS and APM,
and smaller surveys covering less survey area at high redshift. 
It is clear that current clustering data are not very constraining 
on the dark energy equation of state $w_X$, although there seems to be some evidence that
$w_X < -2/3$ may be favored.}
\end{figure}

A number of clustering analyses is presently available for galaxies
at $z\la 5$.
Various factors, such as scale-dependence, type-selection
and Malmquist bias, need, however, to be taken into account to compare the
outcome of different studies (see e.g. Magliocchetti et al. 2000). 
In fact, the clustering properties of galaxies are scale-dependent
and surveys sample a variety of different scales.
Moreover, it is well known that galaxy clustering depends on a series
of characteristics of the galaxy population under scrutiny (e.g.
morphological type, colour, star-formation rate)
and surveys generally use different criteria to select the objects they 
study. Finally, Malmquist bias is 
due to the fact that within a given survey more distant galaxies tend to have 
brighter absolute magnitude and will in general not have the same clustering 
amplitude.
All these effects will have to be taken into account while we compare 
theoretical prediction from various cosmologies with observational data.

Weak lensing surveys have also started to make progress in mapping directly the three
dimensional dark matter distribution in the universe. In the near future such surveys will
not only study the statistical nature of clustering, but will measure the detailed features 
of the underlying mass distribution.
Cross-correlating weak lensing maps with galaxy surveys will provide us with an 
unique way to probe gravitational clustering, and
hence the nature of bias associated with the luminous objects. Therefore it is important to see how 
varying the equation of state in quintessence cosmologies can affect the 
nature of clustering of dark haloes and galaxies. 

The main uncertainty in comparing theoretical predictions about growth of
gravitational instability and observational data of galaxy clustering originates
from the fact that galaxies might be biased tracers of the underlying mass distribution.
In fact,
it is well known that different galaxy populations (selected by morphological type, luminosity,
star-formation rate) cluster differently, hence not all of them can trace the underlying
mass distribution.
A number of models (based on analytical reasoning or numerical simulations) are available to
quantify the expected degree of biasing associated with galaxies and clusters
(see e.g. Magliocchetti \& Porciani 2003 and references therein). 
Most of them associate luminous objects to their hosting dark matter haloes.
A general prediction is that the clustering amplitude of the most massive haloes at any given epoch
is amplified with respect to that of the mass distribution, while very small haloes are nearly 
good tracers of the mass-density field (e.g. 
Mo \& White 1996; Catelan et al. 1998; Porciani et al. 1998, Coles et al. 1999).
Not surprisingly such models are too simplistic
to encompass all the detailed information and the non-linear physics necessary
to understand the formation and clustering of galaxies.
In spite of this, they are able to make reliable predictions of the expected amplitude
of galaxy clustering.
In general, the strong clustering of high-redshift galaxies has been regarded 
as indication of the overall robustness of the theory and as evidence for
the reality of galaxy biasing.

Recent cosmological observations favor an accelerating universe
\cite{Garna98a,Riess98,Perl99}.
This implies the existence of energy of unknown nature (dark energy), 
which has negative pressure. Various observations are consistent with dark energy
being a non-zero cosmological constant (see for example, Wang \& Garnavich 2001;
Bean \& Melchiorri 2002). However, many other alternative
dark energy candidates have been considered, and are consistent with
data as well. For example, quintessence, k-essence, spintessence, etc.
(Freese et al. 1987;
Peebles \& Ratra 1988; Frieman et al. 1995; Caldwell, Dave, \& Steinhardt 1998;
Garnavich et al. 1998b; White 1998; Efstathiou 1999; Steinhardt, Wang, \& Zlatev 1999;
Podariu \& Ratra 2000; Sahni \& Wang 2000; Sahni \& Starobinsky 2000; 
Saini et al. 2000; Waga \& Frieman 2000;
Huterer \& Turner 2001; Ng \& Wiltshire 2001; Podariu, Nugent, \& Ratra 2001;
Sarbu, Rusin, \& Ma 2001; Weller \& Albrecht 2001)

Various dark energy models can be conveniently classified according
to the equation of state of the dark energy component, $w_X$. For
example, for quintessence models, $dw_X /dz > 0$, while for k-essence
models, $dw_X /dz < 0$. However, it is extremely difficult to determine
the time dependence of $w_X(z)$ \cite{Maor01,Barger01,Maor02}.
Wang and Garnavich (2001) have shown that it is more optimal to
constrain the time dependence of the dark energy density $\rho_X(z)$,
instead of $w_X(z)$. In this paper, we only consider
toy models with $w_X=$ constant for simplicity and illustration.
This is appropriate for our purposes, since current galaxy clustering
data can not place useful constraints on the time dependence of
$w_X(z)$.
However, our method can readily be extended to models
with time dependent equation of state. Our results will also have direct 
relevance for programs which focus on reconstructing the potential 
energy $V(\phi)$
of the quintessence field
from observed galaxy clustering data.

There are many other probes of dark energy.
These include, the distance-redshift relations of cosmological standard
candles, Cosmic Microwave Background Anisotropy, volume-redshift test
using galaxy counts, the evolution of galaxy clustering, weak lensing,
etc. These different methods to probe dark energy are complimentary
to each other, and can provide important consistency checks,
due to the different sources of systematics in each method
(for example, see Kujat et al. 2002 and references within).

\section{Evolution of Clustering in Quintessence Cosmologies}

Hamilton et al. (1991) proposed a scaling ansatz for computing the non-linear
matter power spectrum of a given cosmological model at any epoch.
This method was later extended by various authors to reproduce the outcome
of high-resolution numerical simulation in a cold dark matter scenario 
(see e.g. Peacock \& Dodds 1996; Smith et al. 2003). 
In the version by Peacock \& Dodds (1996) that we adopt here, this ansatz 
essentially consists of postulating that 
$4\pi k^3 P(k) = f[ 4 \pi k_l^3 P_l(k_l)]$,
where $P(k)$ is the nonlinear power spectrum and $P_l$ is the linear
power spectrum, and the function $f$ in general will depend on the
initial power spectra. The linear power spectrum is evaluated at a
different wave number, $k_l = [ 1 + 4 \pi k^3 P(k) ]^{-1/3} k$, 
hence the mapping is non-local in nature. 
The form of the function $f$ is calibrated against $N$-body simulations,
by assuming that it matches the predictions of linear theory on large scales,
and of stable clustering on small scales (see Smith et al. 2003 for a critical
discussion).
Ma et al. (1999) showed that, at $z=0$, the Peacock \& Dodds (1996) formula 
is accurate
even in the presence of quintessence. However, at earlier epochs, it tends
to underestimate the non-linear power on scales smaller than $\sim 1 \,h^{-1}$
Mpc by up to 30 per cent. Given that we are interested in galaxy clustering 
on mildly non-linear scales (and given the uncertainties on present-day 
determinations of galaxy clustering at high-$z$), the Peacock \& Dodds 
formulation is good enough for our analysis.

The cosmological model enters the scaling ansatz
primarily through the linear growth function 
$D(z)$\footnote{The matter transfer function has negligible
dependence on dark energy models on the scales of interest to
us \cite{Ma99}.}, so that 
$P_l(k, z) = [D(z)/D(z=0)]^2\, P_l(k, z=0)$.
The linear growth function is evaluated directly from the equation:

\begin{equation}
\ddot D(z) + 2H(z) \dot D(z) - {3 \over 2} H_0^2 \Omega_m (1+z)^3 D(z) =0,
\label{growth}
\end{equation}
where the dots denote derivatives with respect to $t$.
For a constant dark energy equation of state $w_X$, the evolution of the 
Hubble parameter $H(z)$ can be written as:

\begin{equation}
H(z) = H_0[ \Omega_m(1+z)^3 + (1-\Omega_m-\Omega_X) (1+z)^2 + \Omega_X(1+z)^{3(1+w_X)}]^{1/2}\;,
\end{equation}
In general, equation (\ref{growth}) must be solved numerically since
the usual integral equation for $D(z)$ (Heath 1977) does not hold
in the presence of quintessence (unless $w=-1$ or $w=-1/3$).
However, when $\Omega_m+\Omega_X=1$ and $w=$ constant, 
equation (\ref{growth}) can be solved analytically
in terms of hypergeometric functions (Padmanabhan 2003).
Useful approximations for the linear growth functions, 
can be found in Lahav et al. (1991) for 
$\Lambda$CDM and in Wang \& Steinhardt (1998) for QCDM models. See 
Benabed \& Bernardeau (2001) for more on power spectrum evolution in quintessence cosmologies.

In this paper, we assume that the matter density parameter $\Omega_m =0.3$,
the dark energy density parameter $\Omega_X=0.7$, the 
Hubble constant $H_0=h\,100\,$km/s$\,$Mpc$^{-1}$ with $h=0.7$, and the 
rms density fluctuation within a top-hat sphere of radius 8 $h^{-1}$ Mpc
linearly extrapolated to today $\sigma_8=0.8$.
The four quintessence models we have studied are
$w_X = -1/3, 2/3, -1, -1.9$. 
The $w_X=-1.9$ is taken as an example
of the class of models which violate the weak energy condition (WEC) \cite{Wald84} 
of recent theoretical interest \cite{Caldwell02,Frampton02,Onemli02}.

\begin{figure}
\protect\centerline{
\epsfysize = 2.7truein
\epsfbox[25 500 588 715]
{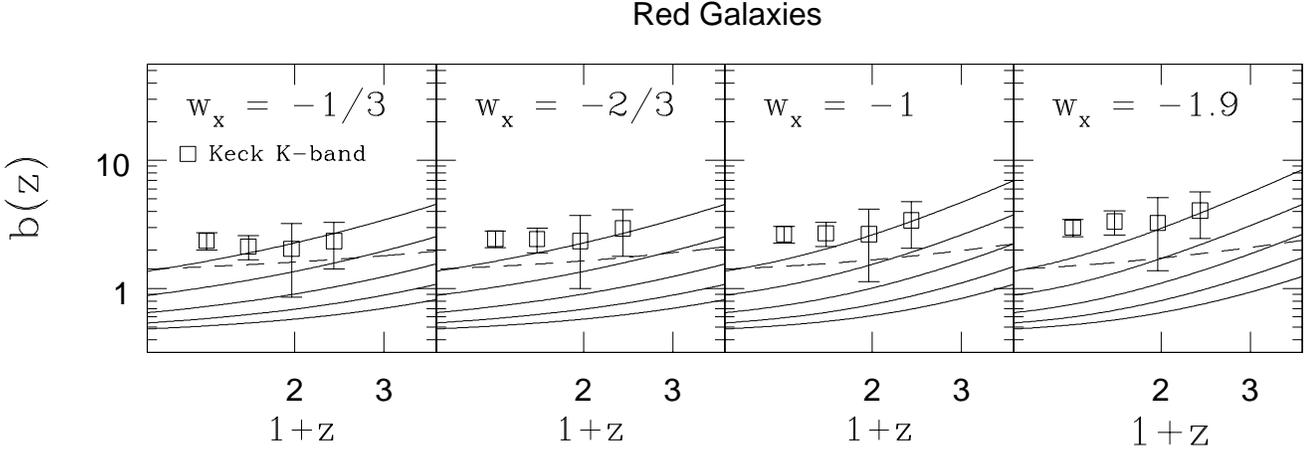} }
\caption{
Analytical computations of the bias parameters for various quintessence models. Dashed lines are results from
test-particle model and solid lines represent computations from the halo model. 
We use the analytical results by Mo \& White (1996) to compute bias for haloes 
larger than a given mass. Curves from bottom upwards correspond to haloes with masses 
greater than $10^9$, $10^{10}$, $10^{11}$, $10^{12}$, and $10^{13}$ M$_{\odot}$. For the
test particle model, we have computed the bias assuming $\sigma_{8} = 1.13$,
as derived from the Stromlo-APM survey (Loveday, Tresse \& Maddox, 1999) for galaxies
with no emission lines (red objects). These bias parameters are computed from the estimated $r_0$ from 
these surveys by Magliocchetti et al. (2000). Note that the bias computed
from the Mo \& White (1996) formalism is not forced to reproduce any observational data.
Observational data points correspond to Keck-K band survey (Carlberg et al. 1997). 
See text for more details.}
\end{figure}

To see how the evolution of clustering varies in quintessence models, we
calculate the rms fluctuation $\sigma_8(z)$ as follows:
\begin{equation}
\sigma_8^2(z) = \int_0^{\infty} \frac{dk}{k} \, \Delta^2_l(k,z)\,
\left( \frac{3j_1(kr)}{kr} \right)^2,
\end{equation}
where $\Delta^2_l(k,z) = 4\pi k^3 P_l(k,z)$, and $r=8\,h^{-1}$Mpc. 
We can similarly define 
the rms fluctuation in galaxy density as $\sigma_8^{(g)}(z)$.
It is sometimes convenient to relate the mass and galaxy fluctuations 
on the $8\,h^{-1}$Mpc scale by
introducing the bias parameter $b_8(z)=\sigma_8^{(g)}(z)/\sigma_8(z)$.

Figure 1 shows analytical computations of $\sigma_8$ 
(and $\sigma_8^{(g)}$) 
as a function of redshift $z$
for various quintessence models, together with observational data. 
Except for four new data points for LBG galaxies
at $z=3$ (Porciani \& Giavalisco 2002; Adelberger et al. 2003),
$z=4$ (Ouchi et al. 2001), and for Ly-$\alpha$ emitters at $z=4.86$ (Ouchi et al. 2003),
the observational data for $\sigma_8^{(g)}(z)$ 
are from
Magliocchetti et al. (2000), converted to each model as described in their 
paper (see Appendix A).
The solid lines
represent the linear growth rate $D(z)$ 
(normalized so that $D\to (1+z)^{-1}$ when $z\to \infty$) 
for various cosmologies as a function of redshift.
The short-dashed lines represent the theoretical $\sigma_8(z)$; 
the two dashed lines are normalized to APM and IRAS surveys at low redshift
respectively. 
The general trend (masked by large error bars) is that $\sigma_8^{(g)}$ 
decreases 
between $z=0$ and
$z=2$ while it either keeps constant or increases at higher redshifts. A similar behaviour is seen
in numerical simulations for the clustering of dark matter haloes (see, e.g., Jenkins et al. 1998).
Note that the most recent data sets correspond to substantially smaller
error bars at high $z$. However, it is important to stress that 
Ouchi et al. (2001, 2003) assumed that the slope of the correlation function is $\gamma=1.8$,
so that the corresponding error bars for $\sigma_8^{(g)}(z)$ 
are under-estimated 
(not including the uncertainty in $\gamma$).

It is clear that current clustering data are not very constraining 
on the dark energy equation of state $w_X$, mainly because the scatter of the data points is large 
in Fig.1. However, since different types of galaxies are
expected to cluster differently, in the next section we will try to reduce this scatter
by dividing the galaxies into subgroups.

\section{Clustering of Galaxies and Dark Matter Haloes in Quintessence Cosmologies}

In general, it is not clear how the spatial distribution of galaxies is related with the underlying 
mass distribution; this relationship 
it is likely to be non-linear, non-local, scale-dependent, type-dependent and
even stochastic (Catelan et al. 1998; Dekel \& Lahav 1999). 
However, due to the lack of the complete picture of how galaxies are formed, 
various analytical and semi-analytical models have been proposed which capture some
basic flavors of galaxy clustering. 

We parameterize the clustering properties of a population of cosmic objects through
a bias parameter $b$ (a function of separation and redshift) defined by the ratio between the galaxy
autocorrelation function, $\xi_{\rm g}$, and the corresponding quantity for the mass density
distribution, $\xi$, as  
\begin{equation}
b^2(r,z)=\xi_{\rm g}(r,z)/\xi(r,z)\;.
\end{equation}
In what follows the scale dependence will be neglected since we will either consider the large
separation limit (for the models) or refer to a limited range of separations accessible to a given
survey (for the data) over which only small variations of the bias parameter are possible.

\begin{figure}
\protect\centerline{
\epsfysize = 2.7truein
\epsfbox[25 500 588 715]
{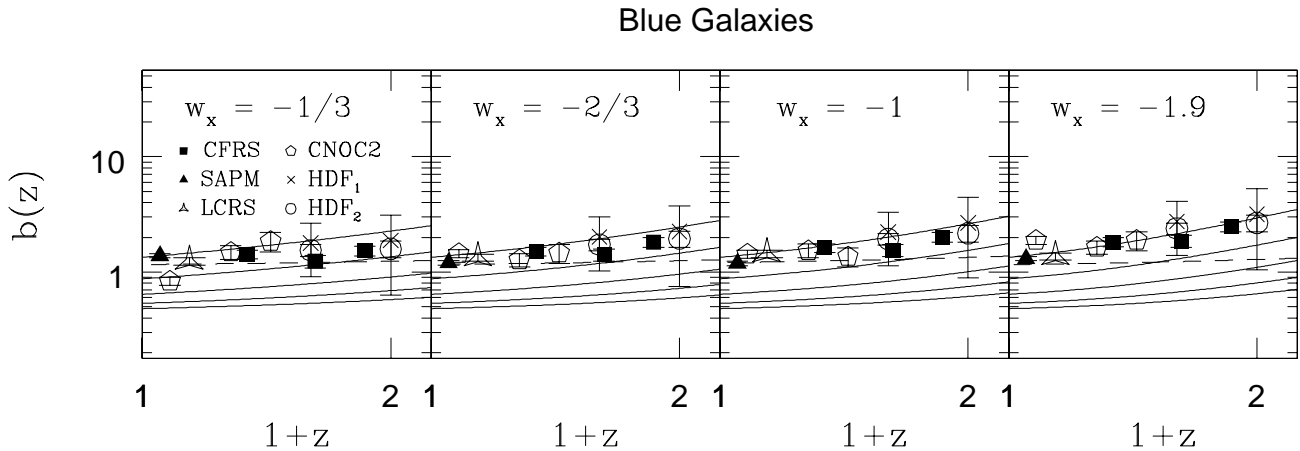} }
\caption{
As in figure 2 but for blue galaxies.
In this case,
for the test particle model we have computed the evolution of the bias parameter assuming $\sigma_8 =0.93$,
as derived for galaxies with weak emission lines 
in the Stromlo-APM survey (Loveday, Tresse \& Maddox, 1999). 
The data points correspond to the following surveys: CFRS 
(Le Fevre et al. 1996), Stromlo-APM (Loveday et al. 1995), LCRS (Huan et al. 1996), 
CNOC2 (Carlberg et al. 1997),  ${\rm HDF}_1$ (Connoly, Szaly \& Brummer, 1998), 
${\rm HDF}_2$ (Magliocchetti \& Maddox 1999).
See the main text for more details.}
\end{figure}

\subsection{No Merger or Galaxy Conserving Model}

We first consider a simple biasing scheme which treats galaxies as
test particles moving in the overall potential generated by the large-scale
structure.
It assumes that a galaxy population is generated at a given cosmic
epoch with a density distribution which is linked to the mass density
by a linear bias parameter.
In other words,
test particles representing galaxies are distributed throughout 
the Universe in such a way that their density contrast is directly 
proportional to the density contrast of the 
underlying mass distribution. This model also assumes that these 
test particles follow the cosmic flow. The conservation of galaxy number 
density then is used to compute the 
evolution of bias associated with these particles.  It can be shown 
that the evolution of this test particle bias can be written as:

\begin{equation}
b(z) = 1 + [b(z_*) - 1]{D(z_*) \over D(z) } = 1 + [b_0 - 1]{D(z=0) 
\over D(z) }\;,
\end{equation}   
where $D(z)$ is the linear growth rate for gravitational clustering 
which typically depends 
on the background dynamics of the Universe, $z_*$ denotes the epoch
of ``galaxy formation'', and $b_0$ is the bias at the
present epoch.
This can be understood as follows. If we assume a certain class of galaxies is formed at 
a particular redshift due to a specific gas-dynamical formation mechanism,
it will carry a specific bias tag, which one can argue is largely independent 
of the local environment and hence constant for a specific galaxy type. 
However, once formed, these galaxies will have to 
move due to the gravitational field. The final expression for the galaxy bias is derived by assuming
constant comoving number density for these galaxies (Dekel 1986; Fry 1986;
Dekel \& Rees 1987; Nusser \& Davis 1994).
This model is also known as the galaxy conserving model (Matarrese et al. 1997). 
However, one should keep in mind that the basic assumption of inert indestructible 
nature of galaxies is not correct. 

In figures (2-4) we plot the test-particle bias parameters 
(dashed lines) associated with various models with quintessence
and compare them against survey results. 
Corresponding values for $\sigma_8$ are displayed in figure-1 (short-dashed lines). 
In figures (2-4) we have divided the observed galaxy population into 3 subsamples.
It is known from earlier studies that various types of galaxies cluster differently. 
Comparing samples which are inherently similar such as red galaxies or
galaxies with strong star-formation rates do tend to reduce the scatter found among 
the clustering properties extracted from different surveys. 

Note that in figure 1 we have compared the observed results against the
theoretical predictions by forcing the galaxy clustering predictions to match low redshift
results from APM and IRAS surveys respectively; on the other hand, in figure 2-4, the theoretical
bias predictions were normalized to different values extracted from various subsamples of
the APM galaxies with similar characteristics. 

All the observational data in Figs.2-4 are from Magliocchetti et al. (2000),
except for the four new data points in Fig.4, which are for LBG galaxies
at $z=3$ (Porciani \& Giavalisco 2002; Adelberger et al. 2003),
$z=4$ (Ouchi et al. 2001), and Ly-$\alpha$ emitters at $z=4.86$ (Ouchi et al. 2003).
Once again we stress that
these four new data points have substantially smaller
error bars. However, Ouchi et al. 2001 and 2003 assumed $\gamma=1.8$,
hence the error bars for $b(\rbar,z)$ are under-estimated (not including
the uncertainty in $\gamma$).

Note that the scatter of the data points in Fig.4 is much larger than
in Figs.2-3, and there are two sets of dashed curves representing
the test particle model in Fig.4.
One set of dashed curves is anchored at low redshift to the Stromlo-APM survey 
(only starburst galaxies), while the other set of dashed curves is
anchored to LBGs at $z=3$.

\subsection{Press-Schechter and Halo Bias}

In order to compute the evolution of galaxy clustering, it is often convenient
to associate galaxies to their host dark matter haloes.
This can be done in many different ways, see e.g. Cooray \& Sheth (2002) 
for a recent review.
In this paper, for simplicity, we will always assume that a given class of cosmic objects corresponds to a halo population with a mass that is above a given
threshold value. The underlying idea is that, at large separations, the
correlation function will be dominated by objects residing in different haloes
and will be similar to the halo correlation function.
The two-point correlation function of dark matter haloes has been the subject 
of 
many recent analytical as well as numerical studies. In particular, the use of 
the peak-background split method (Efstathiou et al. 1988; Cole \& Kaiser 1989)
and
the extended Press-Schechter 
(see e.g. White 2002 for a recent review on Press-Secheter mass function
and related issues) formalism have been combined to compute the correlation 
function of dark matter haloes in Lagrangian space and mapping from Lagrangian space to
Eulerian space within the context of spherical collapse model
(see Catelan et al. 1998 for a more general approach).
Mo \& White (1996) have derived an analytical expression (expected to be valid 
in the large separation limit) for the halo-halo correlation,

\begin{figure}
\protect\centerline{
\epsfysize = 2.7truein
\epsfbox[25 500 588 715]
{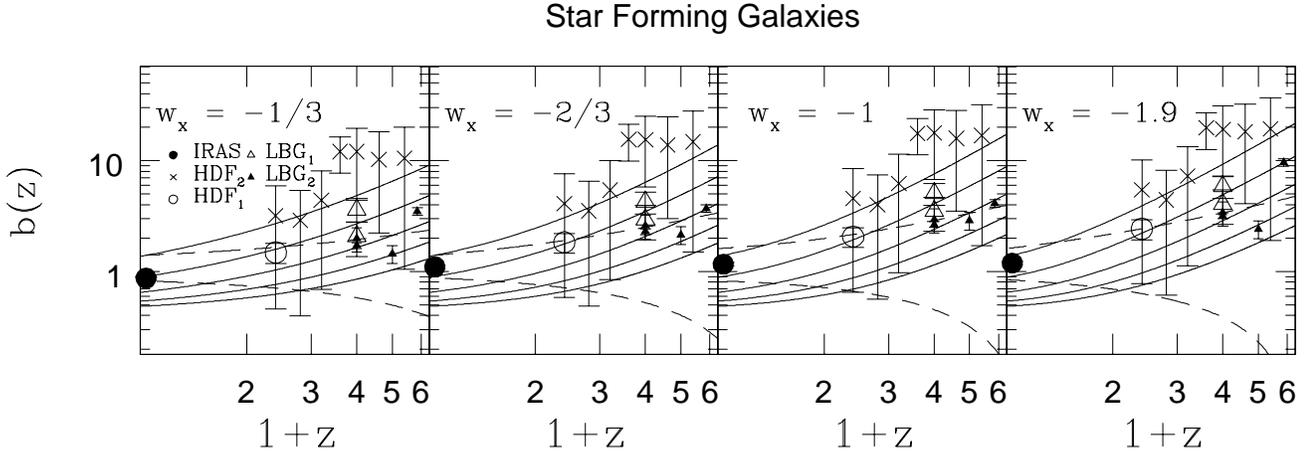} }
\caption{
As in figure 2 but for star-forming galaxies.
In this case, we show two different evolutionary tracks for the bias parameter
in the test particle model. The lower one
is computed assuming $\sigma_8 = 0.66$
as derived from galaxies with very strong emission lines (also classified as 
star forming galxies) in the Stromlo-APM survey (Loveday, Tresse \& Maddox, 1999), while 
the upper one matches the clustering of LBGs at redshift 3.
Observational data points correspond to the following surveys: IRAS (Saunders et al. 1992), $\rm {HDF}_1$ 
(Connoly, Szalay \& Brummer, 1998), $\rm {HDF}_2$ (Magliocchetti \& Maddox 1999) 
LBG$_1$ (Giavalisco et al. 1998, Adelberger et al. 1998),
and LBG$_2$ (Porciani \& Giavalisco 2002; Adelberger et al. 2003; 
Ouchi et al. 2001, 2003).
See the text for more details.}
\end{figure}

\begin{equation}
\xi_{hh}(r; M) = b^2(M) \xi_{mm}(r)\;, 
\end{equation}
where the bias parameter $b(M)$ as computed from the Press-Schechter formalism can be 
written as:

\begin{equation}
b(M) = 1 + {\delta_c \over \sigma^2(M)} - {1 \over \delta_c}\;,
\end{equation}
with $\sigma(M)$ the linearly evolved rms density fluctuation of top-hat spheres
containing an average mass $M$. The parameter $\delta_c$ is derived from the 
dynamics of the spherical collapse in an expanding background. It was shown that the 
parameter $\delta_c$ is largely insensitive to background dynamics of the universe 
(Weinberg \& Kamionkowski 2002). In our studies we have fixed $\delta_c \simeq 1.69$ which
is of sufficient accuracy for our purpose.

Many refinements of the Mo-White calculations can be found in the literature. 
Catelan et al. (1998) followed the non-linear evolution of the clustering
of dark matter haloes using a stochastic approach to biasing.
Jing (1998, 1999) and Porciani, Catelan \& Lacey (1999) showed that an improved
model for halo selection in Lagrangian space
based on sounder theoretical grounds than the naive Press-Schechter 
approach
is required to accurately reproduce the outcome of numerical simulations.
Sheth, Mo \& Torman (1999)
have generalized the formalism by using anisotropic collapse scenarios 
instead of spherical collapse. This model has been calibrated
against N-body simulations in the $\Lambda$CDM cosmology.

It is also possible to construct bias models assuming the hierarchical nature of higher order
correlation functions in gravitational clustering (Bernardeau \& Schaeffer 1989). 
The general trend in such calculations is largely in agreement with halo models (Valageas, Silk
\& Schaeffer 1999). We plan to discuss such models and its relevance in weak lensing surveys
or its cross correlations with galaxy surveys in future publications.

We have coupled the Press-Schechter formalism with the Mo \& White (1996) model to compute the number 
densities (Fig. 5) 
and bias associated with various objects in quintessence cosmologies (Figs. 2-4). 
In order to compare theory and observations, we assume that a given galaxy population corresponds
to observing all haloes beyond a certain threshold or cutoff mass $M_{min}$.
The corresponding clustering properties are then computed by
weighting the bias parameter of haloes of mass $M$ with the appropriate number density. 
Figures 2-4 show the bias parameter for objects heavier than
$10^9-10^{13}$ M$_\odot$. The corresponding values for $\sigma_8$ are also 
plotted in figure 1 (dotted lines).
Our results show a basic degeneracy between the dark energy equation of state and the way
galaxies populate dark matter haloes. Typically we find that objects are 
more biased, and thus correspond to more massive haloes, in cosmologies with more negative 
values of $w_X$.
Hopefully, future surveys will reduce the scatter and the uncertainties of the data-points.
Connecting different populations at different redshifts and understanding the evolution
of the corresponding bias parameters will be crucial to inferring
constraints on dark energy. On the other hand, the degeneracy between cosmology and galaxy 
biasing means that pinning down the biasing scheme
may not be easy until we better understand the properties of dark energy.

Figure 5 shows the number density of haloes as a function of redshift for 
various quintessence models 
versus data points converted to each model from the HDF analysis data of
Magliocchetti \& Maddox (1999).
The different sets of curves correspond to haloes with masses greater than 
$10^9$, $10^{10}$, $10^{11}$, and $10^{12}$ M$_\odot$ from top to bottom.
Different linestyles correspond to different values of $w_X$. The solid line represents the 
$\Lambda$CDM model.
For a given redshift, the data points correspond to decreasing $w_X$ from top to bottom. 
The shape of the theoretical curves is typical of any hierarchical scenario
for structure formation deriving from primordially Gaussian fluctuations. 
At early epochs, the halo number density 
within a given mass interval ($M>M_{\rm min}$) increases with time as 
density peaks of lower and lower amplitude go non-linear 
on the mass scales $M_{\rm min}$. 
The number density then reaches a maximum at the epoch
in which the characteristic mass of the existing haloes coincides with
$M_{\rm min}$, and declines afterwords
when objects in the interesting mass interval merge to form bigger haloes.
Note that $M_{min}$ inferred from galaxy clustering is consistent with their 
abundance at high redshift, suggesting that our simple biasing scheme is 
accurate
enough to describe the basic properties of galaxy clustering.
In facts, it is not surprising that no analytic curve traces the evolution of 
HDF galaxies since selection effects will pick up
totally different populations (probably residing in haloes
with different masses) at low and high-$z$.
Hopefully joint analyses of the evolution of the number density and bias
parameter of different galaxy populations will help shed some light on the viable
cosmological models and biasing schemes.

\begin{figure}
\protect\centerline{
\epsfysize = 3.truein
\epsfbox[21 427 300 715]
{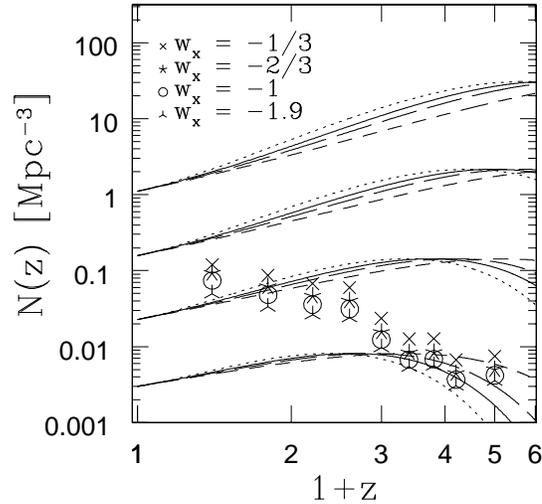} }
\caption{
The number density of haloes for various quintessence models is plotted as a function
of redshift $z$. The different sets of curves correspond to haloes with masses greater than 
$10^9$, $10^{10}$, $10^{11}$, and $10^{12}$ M$_\odot$ from top to bottom.
Different linestyles correspond to different values of $w_X$. The solid line represents the 
$\Lambda$CDM model.
The observational datapoints come from the HDF analysis 
(Magliocchetti \& Maddox 1999).
Note that $M_{min}$ inferred from galaxy clustering
is consistent with their 
abundance at high redshift. The data points correspond to the observed number density 
by assuming a specific equation of state.
For a given redshift, the points correspond to decreasing $w_X$ from top to bottom. }
\end{figure}

\section{Abundance and spatial distribution of galaxy clusters}

The abundance of rich clusters as a function of redshift is a promising tool 
to distinguish cosmological models (Wang \& Steinhardt 1998; Mainini \& Macci\`o 2002).
The key idea is to constrain the amplitude of the power spectrum of density fluctuations 
at intermediate redshifts.

In figure 6 we plot the evolution of the number density of cluster mass haloes
obtained through the Press-Schechter model. It is clear that  
measuring the cluster abundance at $z\ga 1$ could potentially distinguish among different
dark energy models. This can be done by combining cluster data with other observations which 
strongly constrain other cosmological parameters as, for instance, the matter density parameter and
the shape of the linear power spectrum of density fluctuations. 
A simultaneous analysis of the large-scale clustering and the mean abundance of galaxy clusters would 
give tighter constraint on the cosmology (e.g. Schuecker et al. 2003).
In figure 7 we show how the linear bias of galaxy clusters
is expected to evolve with redshift in different dark energy models.
As expected in bottom-up scenarios, rarer objects correspond to a stronger clustering amplitude.
Clearly, the abundance and spatial distribution of galaxy clusters are a sensitive
probe of dark energy at intermediate redshifts.

From the observational point of view, the quest for clusters at intermediate redshifts
is becoming a mature field.
Deep optical and near infrared surveys (which look for local galaxy density enhancements) 
allow the detection of the richest clusters at $z\sim 1$.
Even though spurious detections and selection effects represent serious problems,
these studies will start being suitable for clustering studies
as they cover areas in excess of 100 square degrees
(e.g. the Red-Sequence Cluster Survey, Gladders \& Yee 2000, and 
the Las Campanas Distant Cluster Survey, Nelson et al. 2002).

Alternatively, clusters can be detected in X-rays through the thermal bremsstrahlung emission 
from the hot intracluster plasma. Selection effects in these samples are much easier
to handle with respect to optical surveys. 
A number of {\it ROSAT} surveys easily detected galaxy clusters out to redshifts of $z\sim 0.4$ 
(Ebeling et al. 1996, 1998, 2000, De Grandi et al. 1999; B\"ohringer et al. 2000).
The upcoming XMM Large Scale Structure Survey (Pierre 2000) will provide about 900 clusters out to a redshift
of about 1.
This can provide useful constraints on cosmological
parameters (assuming a tight control on various systematics; see e.g. Refregier et al. 2002).  
Such surveys with uniform sensitivity will 
provide a very useful observational data base to constrain both the number density
and the bias associated with galaxy clusters (see also Moscardini et al. 2000).
Deep multi-colour follow-up programmes can identify and measure the redshift of 
clusters within the range of $0<z<1$, and near infrared observations can supplement 
distant cluster candidates at $z > 1$. Cluster 2-point statistics can be used to
lift the degeneracies involved with estimating the cosmological parameters
by using cluster counts alone. 
Schuecker et al. (2003b) performed a detailed
analysis of 452 X-ray brightest clusters mainly for $z < 0.3$. Cosmological parameter
estimation using the abundance of REFLEX clusters and SNe Ia data can produce powerful
constraints on the equation of state. Such studies should be supplemented by observations of 
clustering of galaxy clusters to enhance their sensitivity to the equation of state.

The Sunyaev-Zel'dovich effect (hereafter SZ), e.g. the upscattering of CMB photons by electrons in the
hot intracluster medium, is another powerful method to detect high-redshift clusters.
For instance the
Massive Cluster Survey (MACS) already detected 8 clusters at $z>0.5$ (La Roque et al. 2003).
A number of future surveys are expected to detect galaxy clusters exploiting the SZ effect 
(see e.g. Weller et al. 2002; Hu 2003).  
Such studies will conduct deep and narrow surveys using interferometric 
arrays as e.g. the Arc-Minute Micro-Kelvin Imager or AMI (Kneissl 2001), the SZ Array (SZA,
Carlstrom et al. 2000) or the Array for Microwave Background Anisotropy 
(AMiBa, Lo et al. 2000). Shallower surveys as the One Centimeter
Receiver Array (OCRA, Browne et al. 2000) will also be useful for their wider sky
coverage. The shallow but nearly all-sky survey 
conducted by PLANCK (whose multi-frequency maps will be used for component analysis)
will be released to the scientific community and can provide a wealth of information
in this direction. For a detailed analysis of 
the clustering properties of galaxy clusters detectable by PLANCK see Moscardini et al. 
(2001). In addition  deep and wide field surveys using 1000 element bolometric
arrays mounted on a telescope at south pole represent other interesting options for 
cluster surveys.
A more rigorous Fisher Matrix analysis 
of error associated with such surveys in estimating various cosmological
parameters and their cross-correlations will be presented elsewhere.

\section{Discussion}

Today we have a concordance that the universe is accelerating, its energy dominated by
dark energy with a strongly negative equation of state. But we know almost nothing of the 
dark energy --- its equation of state $w_X$ or whether this evolves. These two quantities 
hold crucial clues to the underlying fundamental physics. Therefore by mapping the 
expansion history of the universe one can probe the new physics. 
Future distance-redshift observations of
type Ia supernovae (Wang 2000, SNAP\footnote{http://snap.lbl.gov/}) 
should place useful constraints 
on the dark energy density (Wang \& Garnavich 2001; Wang \& Lovelace 2001;
Wang et al. 2003). If these constraints are consistent with a quintessence model, then
one can hope to map the potential associated with the scalar field using
complementary data, including that of galaxy clustering. 
Several new experiments are being carefully designed to probe the dark energy.
Systematic uncertainties rather than merely paucity or imprecision of 
observations
will be the key obstacle; this underscores the critical importance of 
using independent and complementary methods to probe dark energy.

In this paper we have concentrated on 
the effect of the equation state on galaxy clustering.
\begin{figure}
\protect\centerline{
\epsfysize = 2. truein
\epsfbox[21 427 590 715]
{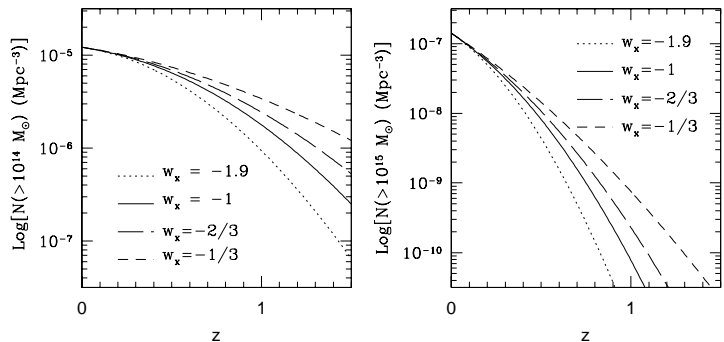} }
\caption{
The number density of cluster-sized haloes is plotted as a function of redshift $z$
for various quintessence models. 
Different curves correspond to different values of the equation-of-state
parameter $w_x$. The left panel correspond to haloes of mass larger than $10^{14} M_\odot$ and the right 
panel correspond to mass greater than  $10^{15} M_\odot$.}
\end{figure}
\begin{figure}
\protect\centerline{
\epsfysize = 2. truein
\epsfbox[21 427 590 715]
{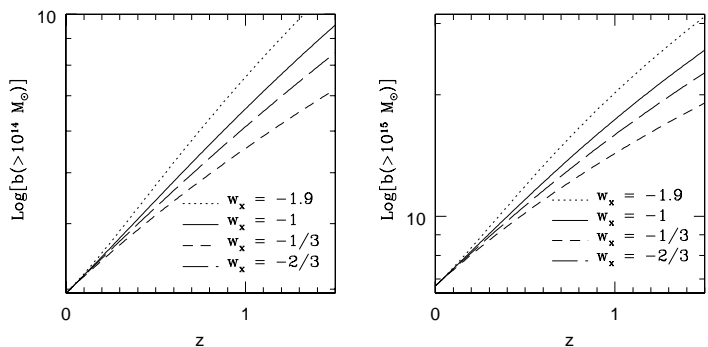} }
\caption{
As in figure 6 but for the bias parameter of cluster-sized haloes.}
\end{figure}
We find that galaxies are more biased (thus corresponding to more massive haloes) 
in models with more negative values of dark energy equation of state $w_X$.
Results from various galaxy surveys are 
corrected of systematic biases and used to compare against quintessence models.
We have shown that correcting the scale dependence of galaxy clustering does reduce 
the observed scatter in estimated bias among various data sets, at least for
moderately high redshifts. 
In spite of this, current data from galaxy clustering do not place strong constraints on
quintessence models (see Figs. 1-4), primarily due to the inhomogeneity of the data
(consisting of many different surveys) and the small area covered by each survey. 
However, our results clearly show the potential of future homogeneous, deep, and
wide-field surveys in constraining dark energy models.
In particular, we have shown that the abundance and spatial distribution of
galaxy clusters from such surveys are 
strongly dependent on the dark energy equation of state 
at intermediate redshifts (see Figs. 6-7).
In future publications, we will study the quantitative constraints
on dark energy that can be derived from future homogeneous, deep, and
wide-field galaxy surveys.

The study of the dynamics of quintessence field directly from observations 
would provide an interesting new independent window to high energy physics. As 
proposed by Starobinsky (1998), luminosity distance measurements of SNe Ia
provides such a possibility. Similarly, 
the observed evolution of clustering of galaxies at 
various
redshifts can be used to construct the potential $V(\phi)$ associated with
the dark energy scalar field $\phi$. 
In principle, one needs to relate the evolution of $ H(z)$ from
the observed evolution of the mass 
density contrast $\delta$ (Starobinsky 1998) 
\begin{equation}
{H^2(z) \over H^2(0) } = { (1+z)^2 \delta'^2 (0) \over \delta'^2(z)}
-3 \Omega_m {(1+z)^2 \over \delta'^2(z)} \int_0^z {\delta(z) |\delta'(z)| \over 1+z} dz\;,
\end{equation}
where primes denote derivative with respect to the redshift $z$.
\footnote{This applies to the linear regime, and corrections are required on smaller scales.}
On large scales, one expects that $\delta$ and fluctuations in the number
density distribution of galaxies, $\delta_g$, are 
related by some bias factor as described above 
$\delta_g(z) = b(z)~\delta(z)$. 
Once galaxy biasing has been specified, 
next one needs to relate the evolution of Hubble parameter $H(z)$ to 
the potential of the scalar field $V(\phi)$  (Saini et al. 2000),

\begin{eqnarray}
&& { {8\pi G \over 3 H_0^2} V(\phi) =  {H^2(z) \over H_0^2} - {(1+z) \over 6H_0^2 } 
{dH^2(z)\over dz} - {1 \over 2} \Omega_m (1+z)^3} \\
&& { {8\pi G \over 3 H_0^2} {\left ( d\phi \over dz \right )^2} = {2 \over 3 H_0^2 (1+z)}
{d {\rm ln} H \over  dz} - {\Omega_m (1+z) \over H^2}  }\;.
\end{eqnarray}
As pointed out before, a simplistic 
linear-biasing picture may not be correct as
we will need a more complete picture of the physics associated with the galaxy
formation process. 
In this paper we have explored a number of analytical models of galaxy bias.
Even though they look plausible when compared to present data, a cleaner 
methodology will probably be required to reconstruct the scalar field potential
directly from galaxy clustering. 
Future weak lensing surveys will be very useful
in this respect and cross correlating weak lensing surveys with redshift 
surveys 
will provide us with a direct handle on $b(z)$ which in turn will be used to
reconstruct the scalar field potential $V(\phi)$. 
However the toy models that we have studied in this 
paper can provide a valuable starting point.


In summary, at present it is not 
realistic to place strong constraints on dark energy from observed galaxy 
clustering.
However, future generation surveys with much higher sky coverage, when 
complemented by 
detailed measurements of evolution of gravitational clustering from weak 
lensing
measurements, will provide direct constraints on evolution of linear growth 
of density perturbations. 
These, combined with the constraints of the dark energy
density from future supernova data (Wang \& Garnavich 2001; Wang \& Lovelace 2001;
Wang et al. 2003), will make it possible not only to constrain but 
perhaps even to reconstruct the potential associated with the scalar field 
$\phi$. 

\section*{Acknowledgments}
DM acknowledges support from PPARC by grant RG28936.
CP has been partially supported by the Zwicky Prize Fellowship program at 
ETH-Z\"urich and by the European Research and Training Network 
``The Physics of the Intergalactic Medium''.
YW acknowledges support from NSF CAREER grant AST-0094335. 
We are grateful to Manuela Magliocchetti
for sharing with us the published data points from various galaxy surveys,
and for helpful discussions. It is a pleasure for DM to acknowledge many
fruitful discussions with members of Cambridge Leverhulme Quantitative Cosmology Group.

\appendix

\section{Converting Observational data for Various Cosmologies}

We assume a power-law form for the 2-point correlation function, $\xi(r,z) = \left [ r / r_0(z) \right ]^{-\gamma}$.
\footnote{
A summary of some observational
results for various surveys can be found in Magliocchetti et al.(1999).}
The transformation among various cosmological models can be derived  by requiring that the angular
correlation function for a given set of galaxies is the same in different cosmologies.
This implies
\begin{equation}
{
r_{02} = \left [  {h_{01} \over h_{02} } \left (  x_1(z) \over x_2(z) \right )^{1 - \gamma}
{E_1(z) \over E_2(z)} \right ]^{1\over \gamma} r_{01}(z), }
\end{equation}
where $E(z)$ is given by:

\begin{equation}
E(z)  \equiv  \sqrt{ \Omega_m(1+z)^3 + \Omega_k(1+z)^2 + \Omega_X \, f(z)}
\label{eq:E(z)}
\end{equation}

\n
 and $x(z)$ is the comoving distance at 
redshift $z$.
Note that our expression for the transformation of $r_0$ between different cosmological
models is equivalent to, but greatly simplified from that of Magliocchetti et al. (2000, 1999).

We have used the expressions given in (Magliocchetti et al. 2000 see e.g.
 Eqs.(5),(6),(8),(17)) to compute $\sigma_8$ and $b^2(r \bar, z)$.
We write
 \begin{equation}
 \sigma_8(\bar{z})= \left\{ \left[ \frac{r_0(\bar{z})}{8} \right]^{\gamma}
 c_{\gamma} \right\}^{1/2}, \hskip 1cm 
 c_{\gamma}= \frac{72}{ (3-\gamma)(4-\gamma)(6-\gamma)\, 2^{\gamma} }.
 \end{equation}
The errors on $r_0$ and $\gamma$ are propagated into the error of $\sigma_8$. 
The scale dependent bias is defined as
\begin{equation}
b^2(\bar{r}, z) = \frac{\xi_g(\bar{r}, z)}{\xi_m(\bar{r}, z)},
\end{equation}
where 
\begin{equation}
\xi_g(\rbar, z) = [\rbar / r_0(z)]^{-\gamma}, 
\hskip 1cm
\xi_m(\rbar, z) = \int \Delta^2(k,z)\, \frac{\sin k\bar{r}}{k \rbar}\,
\frac{ \mbox{d}k}{k},
\end{equation}
with $\Delta^2(k,z)$ denoting the non-linear power spectrum,
calculated using the Peacock \& Dodds 1996 fitting formulae
(normalized to $\sigma_8^{lin}=0.8$ as described above).

For the four new data points we consider (LBG galaxies
at $z=3$, and Ly-$\alpha$ emitters at $z=4.86$), we have followed 
Magliocchetti et al. (2000) in assigning characteristic scales to each survey.
We used $\bar{r}=5\,$h$^{-1}$Mpc for the $z=3$ data point from 
Adelberger et al. 2003. For the other three new data points,
we set the scale $\rbar= \theta_{max}\,x(z)$, and used 
$\theta_{max}$ of $100''$, $16.67'$, and 
$15'$ for the data at 
$z=3$ (Porciani \& Giavalisco 2002), 4 (Ouchi et al.2001), 
and 4.86 (Ouchi et al. 2003) respectively.

\end{document}